\newcommand{\rem}[1]{}
\newcommand{\beq}{\begin{equation}}
\newcommand{\eeq}{\end{equation}}
\newcommand{\beqa}{\begin{eqnarray}}
\newcommand{\eeqa}{\end{eqnarray}}
\newcommand{\ba}{\begin{array}}
\newcommand{\ea}{\end{array}}

\documentclass[pra,twocolumn,showpacs,preprintnumbers,
amsmath,amssymb,floatfix]{revtex4}
\usepackage{hyperref}
\usepackage{epsfig}

\begin{document}

\title{Condensate Fraction of a Fermi Gas in the BCS-BEC Crossover
}

\author{Luca Salasnich}
\author{Nicola Manini}
\affiliation{Dipartimento di Fisica and INFM, Universit\`a di Milano,
Via Celoria 16, 20133 Milano, Italy}
\author{Alberto Parola}
\affiliation{
Dipartimento di Fisica e Matematica and INFM, 
Universit\`a dell'Insubria,
Via Valleggio 11, 22100 Como, Italy}

\date{\today}

\begin{abstract}
We investigate the Bose-Einstein condensation of Fermionic pairs in 
a uniform two-component Fermi gas obtaining an explicit formula 
for the condensate density as a function of the chemical potential 
and the energy gap. We analyze the condensate fraction 
in the crossover from the Bardeen-Cooper-Schrieffer (BCS) state 
of weakly-interacting Cooper pairs to the Bose-Einstein Condensate (BEC) 
of molecular dimers. By using the local density approximation 
we study confined Fermi vapors of alkali-metal atoms 
for which there is experimental evidence 
of condensation also on the BCS side of the Feshbach resonance. 
Our theoretical results are in agreement with these experimental data 
and give the behavior of the condensate on both sides 
of the Feshbach resonance at zero temperature. 
\end{abstract}


\maketitle

\section{Introduction}

Several experimental groups have reported the observation of 
the crossover from the Bardeen-Cooper-Schrieffer (BCS) 
state of weakly bound Fermi pairs 
to the Bose-Einstein condensate (BEC) 
of molecular dimers with ultra-cold two hyperfine component Fermi 
vapors of $^{40}$K atoms \cite{greiner,regal,kinast} 
and $^6$Li atoms \cite{zwierlein,chin}. 
The detection of a finite 
condensed fraction also in the BCS side 
of the crossover \cite{regal,zwierlein}
stimulated a debate over its interpretation
\cite{falco,avdeenkov,diener,pps}.
Extended BCS (EBCS) equations \cite{eagles,leggett,noziers} 
have been used to reproduce in a satisfactory way 
density profiles \cite{perali} and collective 
oscillations \cite{hu} of these Fermi gases. 
As this EBCS mean-field theory is defined for any value of the coupling, it
provides an interpolation between the BCS weak-coupling regime and the BEC
strong-coupling limit \cite{sademelo,engelbrecht}.
Despite well know limitations \cite{sademelo} the EBCS theory 
is considered a reliable approximation for studying the whole 
BCS-BEC crossover at zero temperature, giving a simple and 
coherent description of the crossover in terms of fermionic variables. 

Within the EBCS scheme, we derive an explicit formula for the number of
condensed fermionic pairs in the uniform BCS ground-state.
We use the EBCS equations to study the behavior of this condensate fraction
as a function of the inter-atomic scattering length $a_F$ in the BCS-BEC
crossover: from the BCS regime (small negative $a_F$) crossing the
unitarity limit (infinitely large $a_F$) to the BEC regime (small positive
$a_F$).
In addition, by using the local-density approximation, we calculate the
condensate fraction and density profiles of the Fermi gas in anisotropic
harmonic confinement.
With no fitted parameters, we find a remarkable agreement with recent
experimental results \cite{zwierlein} indicating a relevant fraction of
condensed pairs of $^6$Li atoms also on the BCS side of the Feshbach
resonance.

\section{Extended BCS equations}

The Hamiltonian density of a dilute interacting two spin component Fermi
gas in a box of volume $V$ is given by
\beq 
{\hat {\cal H}} = -{\hbar^2\over 2 m} \sum_{\sigma=\uparrow, \downarrow} 
{\hat \psi}^+_{\sigma}
\nabla^2 {\hat \psi}_{\sigma}
+ g \; {\hat \psi}^+_{\uparrow}
{\hat \psi}^+_{\downarrow}
{\hat \psi}_{\downarrow}
{\hat \psi}_{\uparrow} \; , 
\label{ham} 
\eeq 
where ${\hat \psi}_{\sigma}({\bf r})$ is the field operator 
that destroys a Fermion of spin $\sigma$ 
in the position ${\bf r}$, while ${\hat \psi}_{\sigma}^+({\bf r})$ 
creates a Fermion of spin $\sigma$ in ${\bf r}$. 
The attractive interatomic interaction is modelled by a contact 
pseudo-potential of strength $g$ ($g<0$). 
The field operators satisfy usual anticommutation rules and 
can be expanded in Fourier series, 
${\hat \psi}_{\sigma}({\bf r})= V^{-1/2} \sum_{\bf k} 
e^{i{\bf k}\cdot {\bf r}} {\hat a}_{{\bf k}\sigma}$, in terms of
operators ${\hat a}_{{\bf k}\sigma}$ destroying a Fermion of spin 
$\sigma$ with linear momentum $\hbar {\bf k}$. 
At zero temperature the BCS variational state \cite{bardeen} 
is given by 
$
|\varphi \rangle = \prod_{\bf k} 
\left( u_k + v_k {\hat a}_{{\bf k}\uparrow}^+ 
{\hat a}_{-{\bf k}\downarrow}^+ \right) |0 \rangle 
$, 
where $|0\rangle$ is the vacuum state, $u_k$ and $v_k$ are 
variational amplitudes, and the state $|\varphi \rangle$ 
is normalized to unity if $u_k^2 + v_k^2 =1$. 
The amplitudes $u_k$ and $v_k$ are obtained by imposing 
the minimization of the thermodynamic potential   
\beq 
\Omega = \int d^3{\bf r} \; \langle {\hat {\cal H}}({\bf r}) 
- \mu \; {\hat n}({\bf r}) \rangle  \; , 
\label{ther} 
\eeq 
where ${\hat n}({\bf r}) = \sum_{\sigma=\uparrow, \downarrow} 
{\hat \psi}^+_{\sigma}({\bf r}){\hat \psi}_{\sigma}({\bf r})$ 
is the number density operator and $\mu$ is the chemical potential, 
determined by the condition $N=\int d^3{\bf r}\langle {\hat n}({\bf r})
\rangle$ that fixes the average number $N$ of fermions. 
All averages  
are done over the BCS state $|\varphi \rangle$. 
Accordingly, the standard BCS equation
for the number of particles is
\beq 
N = 2 \sum_{\bf k} v_k^2  \; , 
\label{bcs1} 
\eeq
and for the BCS gap is
\beq
-{1\over g} = {1 \over V} \sum_{\bf k} {1\over 2 E_k} 
\label{bcsGap}
\eeq 
\cite{bardeen,fetter}.
Here
\beq
E_k=\left[(\epsilon_k-\mu )^2 + \Delta^2\right]^{1/2}
\eeq
and
\beq \label{vk}
v_k^2 = {1\over 2} \left( 1 - \frac{\epsilon_k  - \mu }{E_k} \right) 
\, . 
\eeq
with the non-interacting fermion kinetic energy
$\epsilon_k=\hbar^2k^2/(2 m)$.
The chemical potential $\mu$ and the gap energy $\Delta$ are obtained by
solving equations (\ref{bcs1}) and (\ref{bcsGap}).
Unfortunately, in the continuum limit, due to the 
choice of a contact potential, the gap equation diverges in the 
ultraviolet. A suitable regularization is obtained by introducing 
the scattering length $a_F$ via the equation 
\beq
{m \over 4 \pi \hbar^2 a_F} = {1 \over g} + 
{1 \over V} \sum_{\bf k} \frac1{2\epsilon_k} \,,
\eeq 
and than subtracting this equation from 
the gap equation \cite{eagles,leggett,noziers}. 
In this way one obtains the regularized gap equation 
\beq 
-{m \over 4 \pi \hbar^2 a_F} = {1 \over V} 
\sum_{\bf k} \left( {1\over 2 E_k} - \frac1{2\epsilon_k} 
\right) . 
\label{bcs2}  
\eeq 
The EBCS equations (\ref{bcs1}) and (\ref{bcs2}) 
can be used to study the evolution 
from BCS superfluidity of Cooper pairs ($a_F<0$) to the Bose-Einstein 
condensation (BEC) of molecular dimers ($a_F>0$) 
\cite{eagles,leggett,noziers,sademelo}.
This transition is conveniently studied as a function of the dimensionless
inverse interaction parameter $y=(k_{\rm F} a_{\rm F})^{-1}$, where $k_{\rm
F} = (3\pi^2 n)^{1/3}$ is the Fermi wave vector of non-interacting fermions
of the same density \cite{astrakharchik,heiselberg,manini05}.

\section{The condensed fraction}
 
As previously stressed, several properties of ultra-cold Fermi gases 
have been investigated in the last few years by 
using the EBCS equations \cite{perali,hu}. 
Here we analyze the condensate fraction of fermionic pairs 
that is strictly related 
to the off-diagonal long-range order (ODLRO) \cite{penrose} of the system.
As shown by Yang \cite{yang}, the BCS state $|\varphi \rangle$
guarantees the ODLRO of the Fermi gas, namely that, in the limit wherein
both unprimed coordinates approach an infinite distance from the primed
coordinates, the two-body density matrix factorizes as follows:
\beqa\label{2bodydm}
\langle 
{\hat \psi}^+_{\uparrow}({\bf r}_1') 
{\hat \psi}^+_{\downarrow}({\bf r}_2') 
{\hat \psi}_{\downarrow}({\bf r}_1) 
{\hat \psi}_{\uparrow}({\bf r}_2)  
\rangle \\ \nonumber
= \langle 
{\hat \psi}^+_{\uparrow}({\bf r}_1') 
{\hat \psi}^+_{\downarrow}({\bf r}_2')  
\rangle \langle 
{\hat \psi}_{\downarrow}({\bf r}_1)  
{\hat \psi}_{\uparrow}({\bf r}_2)  
\rangle \, . 
\eeqa
The largest eigenvalue $N_0$ of the two-body density matrix (\ref{2bodydm})
gives the number of Fermi pairs in the lowest state, i.e.\ the condensate
number of Fermi pairs \cite{yang,campbell}.
This number is given by 
\beq\label{ODLRO:def}
N_0 = \int d^3{\bf r}_1 \; d^3{\bf r}_2 \; | \langle 
{\hat \psi}_{\downarrow}({\bf r}_1)  
{\hat \psi}_{\uparrow}({\bf r}_2)  
\rangle |^2 ,
\eeq 
and it is straightforward to show \cite{campbell} that 
\beq 
N_0 = \sum_{\bf k} u_k^2 v_k^2 \; . 
\eeq 
It is quite remarkable that such a formula can be 
expressed in a simple form as a function 
of the chemical potential $\mu$ and the energy gap $\Delta$. 
In fact, in the continuum limit
$\sum_{\bf k} \to V/(2\pi)^3 \int d^3{\bf k}
\to V/(2\pi^2) \int k^2 dk
$, taking into account the functional dependence (\ref{vk}) of the
amplitudes $u_k$ and $v_k$ on $\mu$ and $\Delta$, we find
\beq 
n_0 = {N_0\over V} = {m^{3/2} \over 8 \pi \hbar^3} \,
\Delta^{3/2} \sqrt{{\mu\over \Delta}+\sqrt{1+{\mu^2 \over \Delta^2} }} 
\label{con0} 
\; .   
\eeq 
By the same techniques, also the two EBCS equations 
can be written in a more compact form as 
\beq 
-{1\over a_F} = {2 (2m)^{1/2} \over \pi \hbar^3} \,
\Delta^{1/2} \, I_1\!\left({\mu \over \Delta}\right)  \, , 
\label{gbcs1} 
\eeq 
\beq  
n = {N\over V} = {(2m)^{3/2} \over 2 \pi^2 \hbar^3} \,
\Delta^{3/2} \, I_2\!\left({\mu \over \Delta}\right)  \, ,
\label{gbcs2} 
\eeq 
where $I_1(x)$ and $I_2(x)$ are two monotonic
functions which can be expressed in terms of elliptic 
integrals, as shown by Marini, Pistolesi 
and Strinati \cite{marini}. 

\begin{figure}
\centerline{\epsfig{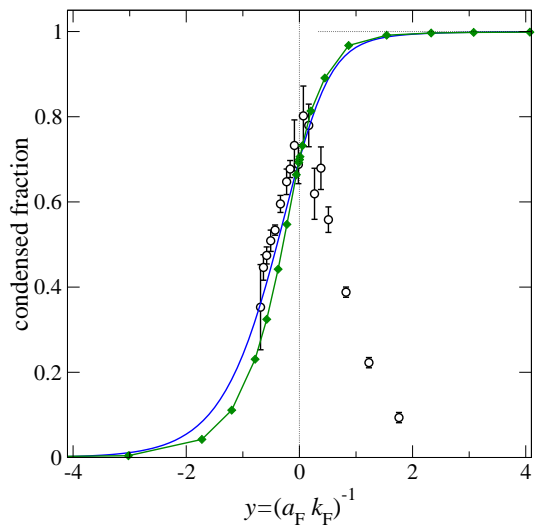}}
\small 
\caption{\label{confrac:fig}
Condensate fraction $N_0/(N/2)$ of Fermi pairs in the uniform two-component
dilute Fermi gas as a function of $y=(k_Fa_F)^{-1}$ (solid line).
The same quantity computed in the LDA for a droplet of $N=6\times10^6$
fermions in an elongated harmonic trap with $\omega_\bot/\omega_z=47$, as in
the experiment of Ref.~\cite{zwierlein}, plotted against the value of $y$
at the center of the trap (joined diamonds).
Open circles with error bars: 
experimentally determined condensed fraction \cite{zwierlein}.
}
\end{figure} 

Simple analytical results from Eqs.~(\ref{con0}-\ref{gbcs2}) 
can be obtained in three limiting cases:
(i) the BCS regime, where $y\ll -1$, $\mu/\Delta \gg 1$ and 
the size of weaklybound Cooper pairs exceeds 
the typical interparticle spacing $k_{\rm F}^{-1}$;
(ii) the unitarity limit, with $y=0$ (thus $a_F=\pm \infty$);
(iii) the BEC regime, where $y\gg 1$, $\mu/\Delta << -1$ and the fermions
condense as a gas of tight dimers.
\\
(i) {\it BCS regime}: $\mu$ approaches the non-interacting Fermi energy
$\epsilon_F = \hbar^2 k_F^2/(2m)$.
One finds that 
$I_1(x) \simeq \sqrt{x} (\ln{(8 x)}-2)$ and 
$I_2(x) \simeq 2 x^{3/2}/3$.
It follows that the condensate density is given by
\beq
n_0 \simeq {m k_F\over 8 \pi \hbar^2} \, \Delta 
= \frac{3\pi}{2 e^2}\, n\, \exp\!\left(\frac{\pi}2\,y\right)\,,
\eeq
with an exponentially small energy gap $\Delta= 8 e^{-2}\, \epsilon_F
\exp{(y\,\pi/2)}$.
\\
(ii) {\it Unitarity limit}: $I_1(x)=0$ and 
$I_2(x)\simeq 1.16$ with $x=\mu/\Delta \simeq 0.85$. 
In addition one finds $\mu \simeq 0.59 \epsilon_F$ and 
$\Delta \simeq 0.69 \epsilon_F$, and the condensate 
fraction is 
\beq\label{confracunitarity}
{2\,N_0\over N} \simeq 0.6994  \,.
\eeq
\\
(iii) {\it BEC regime}: the condensate fraction approaches the ideal Bose
gas value $2\,N_0/N\simeq 1$, with all pairs (molecular dimers)
moving into the condensate.
In addition one finds that 
$I_1(x) \simeq - \pi \sqrt{|x|}/2$ and 
$I_2(x) \simeq \pi/(8\sqrt{|x|})$.
From these expressions 
we obtain 
\beq 
n_0 \simeq {n\over 2} \simeq  {m^2 a_F \over 8 \pi \hbar^4} \Delta^2 \,,
\eeq 
that is precisely the condensate density deduced 
by Pieri and Strinati in this BEC regime 
using a Green-function formalism \cite{pieri2}. 
In this limit the chemical potential
approaches half the dimer binding energy
$\mu \simeq - \hbar^2/(2ma_F^2) = -\epsilon_F \, y^2$
and the gap 
$\Delta\simeq 4\,(3\pi)^{-1/2} \,\epsilon_F \, y^{1/2}$.

The behavior of the condensate fraction $2\,N_0/N$ through the BCS-BEC 
crossover is shown in Fig.~\ref{confrac:fig} (solid line) 
as a function of the Fermi-gas parameter $y$.
The figure shows that a large condensate fraction builds up in the
BCS side already before the unitarity limit, 
and that on the BEC side it rapidly 
converges to the ideal boson gas value.
The finding of a finite nontrivial $2\,N_0/N\neq 1$ around the unitarity
limit contrasts with the suggestion \cite{falco} that the condensate 
fraction could reach unity already at $y=0$.

\section{Fermion gas in a trap}

\begin{figure}[t!]
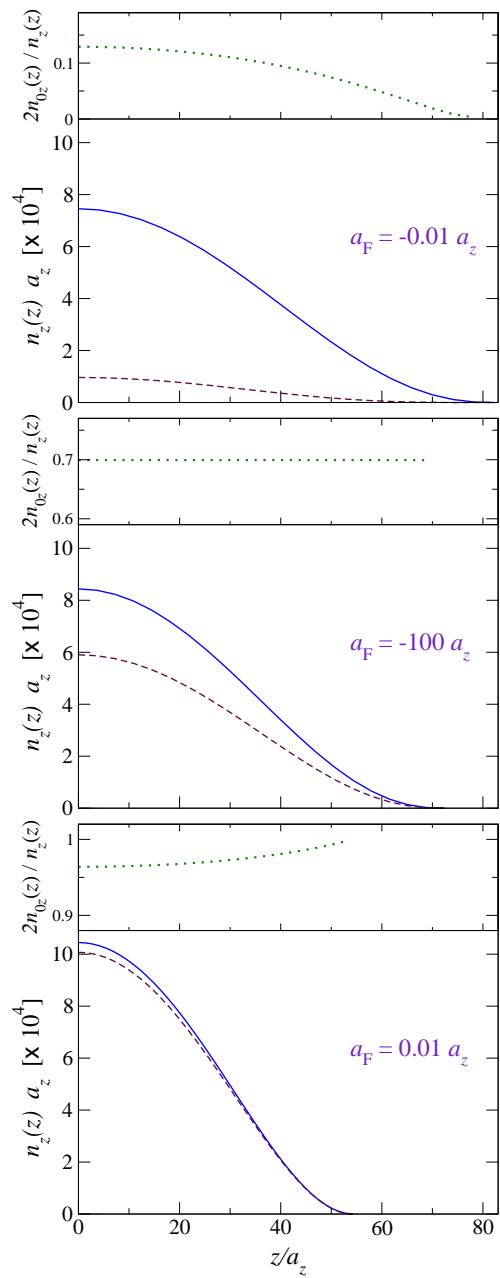

\centerline{\epsfig{file=densProf.-.01.eps,width=6.5cm,clip=}}
\centerline{\epsfig{file=densProf.-100.eps,width=6.5cm,clip=}}
\centerline{\epsfig{file=densProf.01.eps,width=6.5cm,clip=}}
\small 
\caption{\label{densProf:fig}
Axial total density [$n_z(z)$, solid] and condensed density [$2 n_{0z}(z)$, 
dashed] of a droplet composed by $N=6\times10^6$ fermions in an elongated
harmonic trap with $\omega_\bot/\omega_z=47$, as in the experiment of
Ref.~\cite{zwierlein}.
With $^6$Li atoms, the harmonic oscillator lengthscale $a_z=\hbar^{1/2}
(m\omega_z)^{-1/2} =12\,\mu$m in the setup of Ref.~\cite{zwierlein}.
Dotted curves: the axial local condensate fraction $2 n_{0z}(z) / n_z(z)$.
The values of $y$ at the center of the droplet are $-1.2$ ($a_{\rm
F}=-0.01 a_z$), $-10^{-4}$ ($a_{\rm F}=-100 a_z$), and 
$0.87$ ($a_{\rm F}=0.01 a_z$), representative of the BCS, 
unitarity and BEC regimes respectively.
}
\end{figure} 

The local density approximation (LDA) can be implemented to calculate the
properties of a nonuniform Fermi gas of sufficiently slowly varying
density, as in the case of a large number $N$ of particles in a smooth 
trapping potential $U({\bf r})$. 
In the spirit of the Thomas-Fermi method, we replace the chemical potential
$\mu$ with the quantity $\mu - U({\bf r})$ into
Eqs.~(\ref{con0}-\ref{gbcs2}). 
In this way the energy gap $\Delta({\bf r})$, the total density $n({\bf r})$
and the condensate density $n_0({\bf r})$ become local scalar fields.
In order to compare with the experiment of Ref.~\cite{zwierlein} done with 
$^6$Li atoms, we solve the LDA equations for $N=6\times10^6$ fermions
trapped in an external anisotropic harmonic potential well
defined by a transverse frequency $\omega_{\bot}$ 
and an axial frequency $\omega_z$: 
\beq
U({\bf r}) =
\frac m2 \left[\omega_{\bot}^2 (x^2 + y^2) + \omega_z^2 z^2\right].
\eeq
We take an anisotropy ratio $\omega_\bot/\omega_z=47$.
For given $a_F$, and for a given initial guess of $\mu$, the local chemical
potential $\mu - U({\bf r})$ is computed and Eq.~(\ref{gbcs1}) for $\Delta$
is solved numerically, and this calculation is repeated for each ${\bf r}$
on a spatial grid.
The total number $N$ of fermions in the trap is then computed by
integrating Eq.~(\ref{gbcs2}) over the position ${\bf r}$.
The $\mu$ parameter is then adjusted iteratively in order to obtain the
required total number $N$ of fermions within the droplet.
The final density, gap, and $n_0$ functions are determined according to
Eqs.~(\ref{con0}-\ref{gbcs2}).

We find that at the border of the droplet, all three of these functions
vanish together, for both positive and negative $a_F$.
Figure~\ref{densProf:fig} reports the computed axial density
\beq
n_z(z) = \int dx\,dy \, n(x,y,z)
\eeq 
and condensed density $2\,n_{0z}(z)$ (defined analogously) for three values
of the scattering length $a_F$.
The fermionic cloud is rather diffuse in the BCS region, and gets more and
more compact as the BEC regime is approached.
The smooth vanishing behavior of $n_z(z)$ near the cloud border is due to 
the integration over transverse coordinates: $n({\bf r})$ 
instead vanishes with a finite jump in its 
gradient component across the surface.
The same applies for the condensed quantities. Of course, 
this sharp vanishing behavior is a consequence 
of the LDA, while quantum delocalization effects 
(here neglected) should smooth the density profile at the surface 
of the actual droplet. 

As shown in Fig.~\ref{densProf:fig}, 
the condensed fraction (dotted curve) vanishes 
as one moves from center to border of the 
BCS cloud, as pairing acts less efficiently as the density is low.
On the contrary, on the BEC side, the condensed fraction increases to unity
toward the cloud border, as the bosonic gas suffers less from quantum
depletion wherever its density is smaller.
In the unitarity limit, the condensed fraction is in fact constant, and
equals the uniform value of Eq.~(\ref{confracunitarity}): this reflects 
the scale invariance of this special limit of infinitely large scattering
length.

The experiment of Ref.~\cite{zwierlein} 
addresses the problem of measuring the total condensate
fraction of trapped interacting fermions near the
unitarity limit.
Even though it would be hard to demonstrate that the 
measured ``condensed fraction'' coincides 
with the ODLRO $2N_0/N$ defined in Eq.~(\ref{ODLRO:def}), 
we attempt a 
comparison of the computed total condensed fraction, reported in
Fig.~\ref{confrac:fig}, with the experimental data at the lowest
temperature $k_{\rm B} T= 0.05 \epsilon_F$.
In the experiment the scattering length is tuned by means of an external
magnetic tuned across a Feshbach resonance.
Following Ref.~\cite{bartenstein}, the scattering length $a_F$ as a
function of the magnetic field $B$ near the Feshbach resonance is given by
\beq 
a_F = a_b  \left[ 1 + \alpha (B - B_0) \right] 
\left[ 1 + {B_r \over B - B_0 } \right]  \; , 
\eeq 
where $B_0 = 83.4149$ mT, $a_b=-1405\;a_0$, $B_r = 30.0$ mT, 
and $\alpha = 0.0040$ (mT)$^{-1}$. 
The measured condensed fraction at the lowest temperature 
is reported in Fig.~\ref{confrac:fig} as open circles with error bars. 
The general increasing trend and the substantially 
large value of the condensed 
fraction on the BCS side and close to the
unitarity limit agree quite well with the EBCS-LDA calculation 
(joined dimonds). 
On the BEC side the rapid drop of experimental data 
is due to inelastic losses and thus these data are not reliable 
\cite{zwierlein}.  
Note also that the condensate fraction measured near a Feshbach resonance
of $^{40}$K atoms \cite{regal} is much smaller than the values reported in
Ref. \cite{zwierlein}.
The origin of this discrepancy has not been clarified yet
\cite{zwierlein,falco,avdeenkov,diener,pps}.

\section{Discussion}

As suggested in the Introduction, the zero-temperature 
EBCS theory has some limitations.
In particular, it overestimates the bosonic scattering length
$a_B$ between molecular dimers in the BEC regime: the theory predicts
$a_B=2a_F$, where $a_F$ is the two-Fermion scattering length, but Monte
Carlo results \cite{astrakharchik} confirm the four-body scattering
analysis \cite{petrov} which gives $a_B = 0.6 a_F$.
To overcome such difficulties, the EBCS theory can be further extended by
including beyond mean-field corrections which improve the determination of
the chemical potential, and reduce the bosonic scattering lenght 
to $a_B=0.75 a_F$ \cite{pieri}.

A finite-temperature formulation of the EBCS mean-field theory 
is not difficult \cite{fetter,landau}: for the condensed number 
of Fermi pairs we find the formula 
$N_0 = \sum_{\bf k} u_k^2 v_k^2 \; \tanh^2{(\beta E_k/2)}$, 
where $\beta = 1/(k_{\rm B}T)$. It is then straightforward to
express this relation as an explicit function of $\mu$, $\Delta$ and 
$\beta$, pretty much like Eqs.~(\ref{con0}) 
at $T=0$. On the other hand, it is well 
known \cite{noziers,sademelo,engelbrecht,perali} 
that in the BEC regime the finite-temperature mean-field theory 
overestimates  the critical $T_c$. 
This is related to the  dimer breaking energy 
rather than the loss of coherence of the bosonic gas as it should. 
As a consequence, the condensed fraction is overestimated 
near the unitarity limit and in the BEC region: to get reasonable 
results beyond mean-field corrections are needed 
\cite{noziers,sademelo,engelbrecht,perali}.  

In this paper we have shown that the zero-temperature EBCS mean-field 
theory gives a simple and nice formula for the condensed fraction of 
Fermi pairs in the full BCS-BEC crossover. 
On the BCS side of the Feshbach resonance our theoretical results are in 
agreement with the low-temperature measurements of 
the MIT-Harvard collaboration \cite{zwierlein}. On the BEC side 
of the Feshbach resonance the experimental data of Ref. \cite{zwierlein} 
are not reliable due to inelastic losses. In this region our 
predictions may be a guide for further experimental investigations. 

\section*{Acknowledgments} 

The authors acknowledge D. Pini and L. Reatto for enlightening discussions,
and C.E. Campbell for useful e-suggestions.

\end{document}